\documentstyle[aps]{revtex}

\begin{document}
\title {Searching for Granularity of the Vacuum \protect\\ }
%Phys.Rev.Lett. {\bf 00}, 0000 (1995)}
\author{D.Eichler   
{\it Ben-Gurion University,
Beer-Sheva,  84105, Israel}} 

\maketitle
\bigskip

{\small\begin{quotation}
The hypothesis is considered that  the vacuum is a Lorentz 
non-invariant foam in which translational 
symmetry is spontaneously broken
at the  Planck or grand unification scale. This could possibly be 
observed via  Rayleigh 
scattering of ultrahigh energy quanta, and it appears to be ruled out for
spin 1/2 leptons and independently testable for photons. A weaker version of 
the hypothesis
predicts (for otherwise massless neutrinos)  irreversible neutrino mixing 
over a length 
scale of $l_{Planck}( E_{\nu}/E_{Planck})^{-2}$ in  comoving cosmic 
coordinates  for a purely four dimensional universe, and, 
for neutrinos that  have spin 3/2 in a higher dimensional manifold 
above the energy scale $\eta E_{Planck}$, the mixing is predicted to 
take place over a 
scale of $\eta^{-4}$ times this 
length. Such neutrino mixing might also be observable with atmospheric 
and solar neutrino experiments. \end{quotation}}
 \bigskip\bigskip

\section{Introduction}
Translational invariance of the vacuum  is a sacred 
tenet of physics. 
One might suspect that, near  the Planck scale, space may be grainy 
due, say,
to condensates, 
quantum black holes, a spatially varying $\theta$ parameter, or phase 
decorrelation
of virtual modes.  
 This raises 
the question of whether on such small scales translational symmetry is 
preserved or spontaneously broken. A grainy vacuum would suggest 
that 
 physics at macroscopic scales is to some (possibly small) degree 
unpredictable. Hawking $^{1,2}$ has suggested that the formation and 
evaporation of quantum black holes  leads to such unpredictability.
 The question 
has also been raised as  to whether 
wavefunction collapse in quantum 
mechanics, an apparent breakdown of determinism,  
has anything to do with strong non-linear effects in quantum gravity, 
though 
no self-consistent picture has been constructed along these lines.
  More generally, any Lagrangian  that has non-renormalizable 
terms  that contain field derivatives of higher order than the 
usual kinetic term $\partial_{\mu}\phi\partial ^{\mu}\phi$ may
imply breaking of translational symmetry.

Probably, any breaking of translational invariance also involves 
breaking of Lorentz invariance.  This is because otherwise there would be 
energy non-conservation, which, if due to Planck scale effects, would be 
unacceptably conspicuous. On the other hand, a Lorentz invariant vacuum that 
contains quantum black holes could contain an 
infinite density of them because of the infinitude of Lorentz boosts, so, 
if they were to have any physical effect on objects of arbitrarily high 
energy, then a high energy regularization procedure would surely be a 
prerequisite for describing such an effect, and such a procedure might
discard Lorentz invariance 
at sufficiently high energy.

It is pointed out here  that  observational limits on Rayleigh scattering 
of spin 1/2 particles  rule out a Lorentz non-invariant grainy vacuum. 
Although the discussion is hampered by the lack of 
a specific theory of quantum gravity,  we 
make a simple model in which the conclusions are quantitatively 
robust.

  The results of the Einstein-Padolsky-Rosen  experiment tell us that
nature can coordinate     phases even over space-like separations
when it is necessary to do so  to protect  the most basic kinematical
conservation laws.  But this result would not be obvious without the
experimental results.
Given the logical possibility  that quantum gravity
may  be more complicated than already existing theories,
which could make it "harder" to coordinate phases, there is
{\it a priori} motivation for
doing other EPR-type experiments.  Looking for Rayleigh scattering 
of extremely high energy quanta that traverse large distances
is proposed in this spirit.

\section {\bf Spontaneous Breaking of Translational Invariance}
It is assumed here that the interaction between 
a quantum and "grain" of vacuum  preserves unitarity and 
merely changes the phase of wavefunctions in some 4 + n 
dimensional space that includes the four external space time dimensions 
and internal dimensions as well. (Possibly, one could 
attribute the observed local gauge symmetries in nature to the 
grains themselves, for any  low energy physics 
that survives their effects must be robust to local changes of phase.) The 
effects of the grains are assumed to be local to a very good 
approximation so that,
although the wavefront is corrugated by the grains, the angular 
momentum  relative to  any particular point  is conserved by the  
interaction with a grain at that same  point. 

   One might suppose that the particular mechanism of Planck scale 
quantum black holes would be hard to reconcile with Lorentz invariance at 
all scales. If Lorentz invariance 
is an exact symmetry of the vacuum, then there are an infinite 
number of quantum black holes at any point, corresponding to the infinity 
of velocity space at each  spatial coordinate. The ability of  QBH's to 
swallow an object would not decrease with relative velocity, so their 
opacities would add up.  Moreover, the noise generated by the quantum black 
holes   gives rise to a sort of
Olber's paradox: Any point at any time is subjected to the output of 
 black hole evaporation at all other points within its backward light 
cone.  Hence any regularization procedure that deals with this difficulty 
might have to include some sort of cosmological assumption, and define 
a select frame of  reference.
There is some literature on the  possibility that Lorentz invariance is 
merely a low energy symmetry$^4$.  There may, however, be some differences 
between those proposals and what is considered here.  In any case,  
 the dimensional argument given below 
does not  make sense if Lorentz invariance is exact at arbitrarily 
high energies. 

  In this paper we  merely assume that 
energies and hence wavelengths $\lambda$  are defined in  comoving 
cosmological coordinates, and that 
the sum of the effects of all scattering grains on any point in space time 
can 
be roughly represented by a spherical scatterer with a scattering length 
of order the Planck length. %(Olber's paradox would in any case change 
%this by only a logarithmic factor of order $10^2)$.
   
  Consider, then, the  scattering of a massless particle by a 
more or 
less closely packed "foam" of individual grains (e.g. quantum black holes)  
and let us model the 
grains  as something like packed spheres of radius $l_P\sim 10^{-33}$cm. 
The scattering of a massless particle of spin s requires a finite 
amplitude for a helicity  flip of
2s, but conservation of angular momentum relative to the cell
  that scatters 
thus requires a change in orbital angular 
momentum  of 2s.  The scattering cross section is thus of order 
$l_P^2 (l_P/\lambda)^{4s}$. 
 
 Consider photons: s=1.  In traversing  cosmological 
distances,  
a photon traverses of order $10^{60} l_P$, so the universe  might 
plausibly 
become opaque to quanta above $\sim10^{-15}m_P\sim 10^{13}$eV. 
Curiously, although photons just below this energy have been detected from 
point sources, 
there is no firm evidence that any high energy quanta above this energy 
propagate freely as they are all charged particles that are in any case 
deflected by cosmic magnetic fields. 
(But see reference 5.)

Below $10^{13}$ eV, although the optical depth  to large angle Rayleigh 
scattering  is in any case less than unity, there may still be small angle 
refractive scattering.
Consider first  the limit on deviations from Newton's first law that comes  
from
ultrahigh energy astronomy, which at present can detect TeV gamma rays
from  distances of   order 100 Mpc  (Markarian 501) with a
variability  timescale of order $10^3$ seconds.$^6$  This implies that the 
angular deflection   is less than  $10^{-6.5}$. 

 If one assumes  that the index of  refraction of the vacuum  is
significantly affected by the presence of Planck scale 
foam, that is, the presence of the foam  has a 
significant effect on the propagation of light relative to 
the "bare" vacuum, then fluctuations in the index of refraction 
over a scale of order wavelength $\lambda$ are of order 
$(\lambda/l_P)^{-3/2}$.
Over one wavelength $\lambda$, a photon is "wing" Rayleigh scattered by  an  
angle of order $(\lambda/l_P)^{-3/2}$. In traversing a cosmological
distance D, the  photon is scattered of order $(D/\lambda)$ times,
yielding a net  scattering angle of order   $l_P^{3/2}D^{1/2}/\lambda^2$. 
For Tev photons ($\lambda\sim 10^{-16}$ cm),  this is about $10^{-4}$, 
somewhat in
excess of the observed limits.

Now consider spin 1/2 particles. Here the scattering cross section 
scales as $l_P^2 (l_P/\lambda)^2$. Identifying neutrinos as massless 
spin 1/2 particles implies a cosmic mean free path for $10$ MeV 
neutrinos  of only $10^{9}$cm.  This is difficult to reconcile 
with solar neutrino data and with the supernova 1987A neutrinos. 
Similar problems exist for electrons given the survival of high energy 
storage beams in accelerators.

If the familiar spin 1/2 fermions are in fact spin 3/2 gravitinos 
in the 4+n dimensional space$^7$  one must consider that the 
scattering grains  are small compared to the  
compactified dimensions. In this case, the scattering cross section 
scales 
as $(l_P/\lambda)^6$ on scales smaller than the compactification scale.  
Thus, for a compactification scale that is $\eta^{-1} l_P$, the mean 
free path of the "familiar spin 1/2 " particle is increased by
$\eta^{-4}$, up to $10^{17}$ cm at $E_{\nu}=10 MeV$ for  $\eta$ 
as
low as 1/100.
This is still ruled out by SN 1987A.

Note that a Lorentz invariant version of translational symmetry breaking 
would necessarily imply energy non-conservation and could be tested on 
essentially 
every electron, taking the energy of the electron to be its rest mass. 
The existence of stable atoms over a Hubble time appears to immediately 
rule out this possibility.

\section {\bf Neutrino Mixing}  
We now consider the possibility that  the phases of the grains
are coordinated such that linear momentum and energy  are conserved as 
well as 
charge and angular momentum.  In this case, interactions with the 
grains (e.g. absorption and reemission by  quantum 
black holes) can alter only the  non-conserved quantum numbers of a 
particle 
such as the  generation type of a massless neutrino. This is more or less 
equivalent to representing the grains as internal lines, but possibly 
with some loss of phase information in their propagators. If neutrinos 
are massless (except for the interaction with the grains) they would in the 
above picture simply be 
scattered among different generations so that the neutrino flux of any 
type emitted by a source would eventually approach the average over the 3
generations. This 
would be similar to oscillations except that the  change in 
neutrino type would be irreversible, and the  mass matrix would have
a random component with imaginary expected eigenvalues. 
Applying the simple dimensional arguments discussed above and assuming 
neutrinos to be spin 1/2 down to the Planck scale, the 
mean free path for a 30 GeV neutrino would be about 10$^{2}$ cm. 
This is apparently ruled out by  post-beam dump experiments at 
large accelerators, which set a minimum oscillation length of 
order a km for 30 Gev neutrinos. 

If the neutrinos have spin 3/2  in a higher dimensional manifold that 
has dimensions $\eta^{-1}$ times the Planck length,
 then   the same dimensional arguments raise the 
neutrino mixing length by  $\eta^{-4}$. If $\eta \approx 10^{-3/4}$, then the mixing 
length for 30 GeV neutrinos is of order 10$^{5}$ cm,
probably allowed by present experiments but  testable with  
next generation experiments. 

The range of possible  values for $\eta$  still has enough
flexibility that, together with the  rough nature of the
arguments,  accurate quantitative predictions are not attempted here. 
The predicted level of atmospheric neutrino mixing could be 
detectable for reasonable values of $\eta$, as the accelerator experiments
allow an $E^{-2}$ mixing length to be less than an 
Earth diameter for E $\geq$ 0.5 Gev (and implying in the present context 
a compactification length scale within an order of magnitude or so of the 
Planck scale),
but an {\it a priori} 
quantitative prediction is not attempted here. 
We note, however, that if atmospheric neutrinos are produced in the ratio 
of two muon types, no tau types to one electron type, the post-mixing
distribution is equal numbers of each, as if the muon neutrinos had
oscillated with tau neutrinos.  Differences between the two hypotheses
would arise, however, in the ratio of neutrinos to anti-neutrinos, and in
the energy dependence of the mixing  length.

Finally, the hypothesis of neutrinos mixing over a length 
that is proportional to $E^{-2}$ implies that if atmospheric neutrinos are 
mixed over a length scale of order 0.1 Earth radius, then solar Boron 
neutrinos 
($E\sim 10$ MeV) are mixed over a scale of less than 1 astronomical unit,
 whereas the pp neutrinos ($E\sim 0.26$ MeV) have a range that exceeds an 
A.U. Thus, the Boron neutrinos flux at Earth would be reduced by a factor 
of 3, whereas the pp neutrino flux at Earth would be nearly unchanged.
 A quantitative fit to the GALLEX, Homestake, and SuperKamiokande data 
is made in a companion paper.$^8$

I am grateful to Y. Avishai, R. Brustein, M. Berry,  J. Bekenstein, Y. 
Band, A. Dar, E. Gedalin, E. Guendelman, S. Nussinov, D. Owen, Z.Seidov  
and V. Usov 
for vital discussions. I acknowledge support from the  Israel-U.S.  BSF.
 \begin{itemize}
\item[[1]] Hawking, S.W., Nature, {\bf 248}, 30 (1974).
\item[[2]] Hawking, S.W., Commun. Math. Phys., {\bf 43}, 199 (1975).
\item[[3]] Unruh, W.G., Phys. Rev., {\bf 14}, 870 (1976).
\item[[4]]"Origin of Symmetries" ed. C.D. Froggat and H.B. Nielsen 
(World Scientific, 1991).
\item[[5]] Milgrom, M.,   Usov, V., ApJ. Lett., {\bf 449}, L3-7 (1995).
\item[[6]] Lorenz, E., talk presented at MAGIC Workshop, Barcelona, May 
1997.
\item[[7]] "Introduction to Supersymmetry" by Freund, P.G.O., (Cambridge Monographs on 
Mathematical Physics, 1986).
\item[[8]] Eichler, D., and Seidov Z. Physics Letters, (submitted).
\end{itemize}

\end{document}